\documentclass[twocolumn]{aastex631}

\DeclareRobustCommand{\textsc}[1]{{\mdseries\scshape #1}}
\providecommand{\uat}[2]{\href{http://astrothesaurus.org/uat/#2}{#1 (#2)}}
\makeatletter
\@ifundefined{@keys@name}{}{%
  \def\@keys@name{\textit{Unified Astronomy Thesaurus concepts:}\/~\mbox{}}}
\@ifundefined{frontmatter@keys@format}{}{%
  \g@addto@macro\frontmatter@keys@format{\hangindent=0pt\hangafter=0\relax}}
\makeatother

\makeatletter
\def\sec@upcase#1{#1}
\def\abstractname{Abstract}
\long\def\@makecaption#1#2{\vskip\abovecaptionskip{\normalsize{\bf #1.} #2\par}\vskip\belowcaptionskip}
\def\fnum@figure{Figure~\thefigure}
\def\fnum@table{Table~\thetable}
\makeatother
\usepackage{etoolbox}
\patchcmd{\thebibliography}{REFERENCES}{References}{}{\PackageWarning{main}{References heading patch failed}}

\usepackage{amsmath,amstext,mathrsfs}
\usepackage{bm}

\defcitealias{LazarianPogosyan2016}{LP16}
\defcitealias{LazarianPogosyan2012}{LP12}
\defcitealias{Lazarian2026}{LPM26}
\defcitealias{Melnichenka2026}{MLP26}
\newcommand{\LP}{\citetalias{LazarianPogosyan2016}}
\newcommand{\LPM}{\citetalias{Lazarian2026}}
\newcommand{\MLP}{\citetalias{Melnichenka2026}}
\newcommand{\Lb}{L_{\bar\phi}}
\newcommand{\kp}{k_\parallel}
\newcommand{\avg}[1]{\left\langle #1\right\rangle}

\graphicspath{{./}{fig/}}

\shorttitle{Polarization Directions with Coincident Emission and Faraday Rotation}
\shortauthors{Melnichenka, Lazarian, \& Pogosyan}

\begin{document}

\title{Polarization Directions in Turbulence with Coincident Emission and Faraday Rotation}

\author[0009-0004-1969-779X]{A. Melnichenka}
\affiliation{Department of Physics, Berea College, Berea, KY 40404, USA}

\author[0000-0002-7336-6674]{A. Lazarian}
\affiliation{Department of Astronomy, University of Wisconsin--Madison, Madison, WI 53706, USA}

\author[0000-0002-7998-6823]{D. Pogosyan}
\affiliation{Department of Physics, University of Alberta, Edmonton, AB T6G 2E1, Canada}

\begin{abstract}

Synchrotron polarization carries information about both the magnetized turbulence that emits it and the turbulent plasma that Faraday-rotates it. The polarization direction, normalized to unit amplitude, is measured from Stokes maps at a single frequency, and its spatial spectrum, the directional spectrum, recovers the rotation measure spectrum of a foreground Faraday screen. In the interstellar medium, however, synchrotron emission and Faraday rotation often arise in the same volume, and radiation from different depths is rotated by different angles before it adds up. We test the recent theory of directional statistics for this case with synthetic observations of MHD turbulence and of Gaussian random fields. Weak Faraday rotation adds to the directional spectrum a term with the slope of the rotation measure spectrum and about one third of the screen amplitude, and in MHD turbulence this term dominates the small scales. Under strong turbulent rotation the predictions agree with the simulations only when the rotation measure fluctuations keep their actual, intermittent statistics, while a Gaussian model decorrelates the directions far too much. Under strong mean rotation the wavelength selects one line-of-sight Fourier mode of the emitting turbulence. Single-band polarization surveys, such as those of LOFAR and the SKA, can therefore measure both the spectrum and the intermittency of the magneto-ionic medium.

\end{abstract}


\section{Introduction}
\label{sec:intro}

Turbulence is ubiquitous in magnetized astrophysical plasmas \citep{Elmegreen2004, Brandenburg2013}, and synchrotron polarization carries its imprint over large volumes of the interstellar medium and of other galaxies \citep{Beck2015, Haverkorn2015, Akahori2018}. The polarized emission records the magnetic field in the plane of the sky, and Faraday rotation records the field along the line of sight (LOS) weighted by the density of thermal electrons. \citet{LazarianPogosyan2012} and \citet[][hereafter \LP]{LazarianPogosyan2016} related the correlations of the synchrotron intensity and polarization to the spectra of the magnetic field and of the Faraday rotation fluctuations, and synthetic observations of turbulence have confirmed these relations \citep{Herron2016, Lee2016, Zhang2016, Zhang2018, Xiao2025}.

\begin{figure*}[t]
\centering
\includegraphics[width=0.98\textwidth]{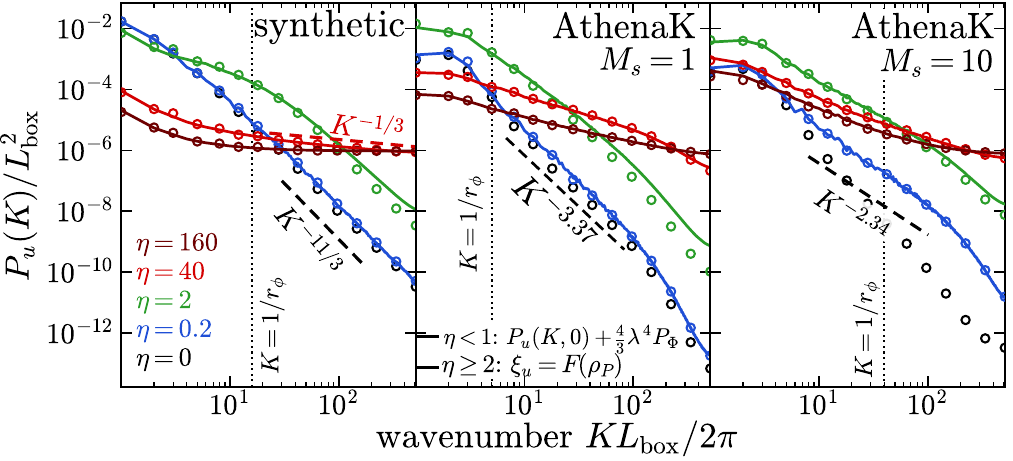}
\caption{Directional spectrum of the two MHD cubes (LOS $\perp\bm B_0$) and of the synthetic cube with the turbulent source, from no Faraday rotation ($\eta=0$) to strong rotation ($\eta=160$), with the colors labeled in the left panel. Markers are measured on the Stokes maps, and lines are the predictions written in the middle panel. For weak rotation, $\eta<1$, the prediction is the spectrum without rotation plus the Faraday term $(4/3)\lambda^4P_\Phi(K)$, Equations~(\ref{eq:weakspec}) and (\ref{eq:fW}). For strong rotation, $\eta\geq2$, it is the spectrum of $\xi_u=\mathcal{F}(\rho_P)$, Equation~(\ref{eq:closure}), with the coherence $\rho_P$ from the measured RMs of the cube for the MHD cubes, Equation~(\ref{eq:xiPmeas}), and from a Gaussian RM for the synthetic cube, Equation~(\ref{eq:xiP}). Lower dashed segments show the slope $K^{-3-m_\phi}$ of the RM spectrum, and the upper dashed segment in the synthetic panel the limiting slope $K^{-1+m_\phi}$ of strong turbulent rotation, Equation~(\ref{eq:inner}), drawn from $K_a$ to the resolution limit. Wavenumbers are in units of the box, and the dotted line marks $K=1/r_\phi$.}
\label{fig:overview}
\end{figure*}

\citet[][hereafter \MLP]{Melnichenka2026} introduced a statistic that uses only the direction of the polarization. The polarization at each position is divided by its amplitude, and the spatial power spectrum of this unit direction field, the directional spectrum, is computed from one pair of Stokes maps at a single frequency. Bright and faint regions enter with equal weight, so the directions respond strongly to Faraday rotation, and for a Faraday screen in front of the emission the directional spectrum recovers the spectrum of the rotation measure (RM) of the screen.

In much of the diffuse Galactic emission, however, the regions of synchrotron emission and Faraday rotation coincide \citep{Iacobelli2013, Jelic2015, Lenc2016, VanEck2017, Wolleben2021}. Radiation from different depths then reaches the observer with different rotation angles and partly cancels \citep{Burn1966, Sokoloff1998}, which shapes the wavelength dependence of the polarized intensity \citep{Arshakian2011, Horellou2014} and the Faraday depth spectra of RM synthesis \citep{Brentjens2005, Ideguchi2014, Basu2019}. Because the polarization is normalized only after the emission of all depths is summed, the screen result does not apply. \citet[][hereafter \LPM]{Lazarian2026} developed the theory of the directional correlation and spectrum for coincident regions, for weak Faraday rotation, for strong rotation by turbulent fluctuations of the RM density and for strong rotation by its mean.

The purpose of this paper is to test this theory with synthetic observations of sub-Alfv\'enic MHD simulations, one transonic and one supersonic, and of Gaussian random fields, in which every element of the volume both emits and rotates. We ask whether weak rotation adds the predicted Faraday term to the directional spectrum, whether the predictions for strong turbulent rotation, derived for Gaussian RM fluctuations, hold for the RM of MHD turbulence, and whether strong mean rotation selects a LOS Fourier mode of the emission. Each prediction is computed from the three-dimensional statistics of the same field, without free parameters. The synthetic cubes satisfy the assumption of the theory that the RM fluctuations are Gaussian and independent of the emission, and the MHD simulations add the non-Gaussian statistics of real turbulence. Figure~\ref{fig:overview} shows how the directional spectrum of the simulations changes from weak to strong rotation. The normalized gradient of the polarization \citep{Gaensler2011, LazarianYuen2018, Herron2018} is the subject of a companion paper.

\section{Theoretical Predictions}
\label{sec:theory}

\subsection{Polarization and Its Direction}
\label{subsec:observables}

Following \LP, each element of the volume emits radiation with the intrinsic polarized emissivity $P_i=Q_i+iU_i$, the polarized intensity density of \LP, and rotates the radiation that passes through it. The polarization observed at the sky position $\bm X$ and wavelength $\lambda$ from a volume of depth $L$ is
\begin{equation}
P(\bm X,\lambda^2)=\int_0^L dz\,P_i(\bm X,z)\,e^{2i\lambda^2\Phi(\bm X,z)}
\label{eq:Pmixed}
\end{equation}
where the Faraday RM between the observer and the emitting depth $z$ is
\begin{equation}
\Phi(\bm X,z)=\int_0^z ds\,\phi(\bm X,s)
\label{eq:Phi}
\end{equation}
and $\phi\propto n_eB_\parallel$ is the RM density, proportional to the thermal electron density and the LOS magnetic field. The RM of the whole depth, $\Phi(\bm X,L)$, forms the RM map.

The polarization direction is the polarization normalized to unit amplitude,
\begin{equation}
u(\bm X,\lambda^2)=\frac{P}{|P|}=e^{2i\chi(\bm X,\lambda^2)}
\label{eq:u}
\end{equation}
where $\chi$ is the polarization angle. The directional correlation of two lines of sight separated by $\bm R$ on the sky is
\begin{equation}
\xi_u(\bm R)=\avg{u(\bm X)\,u^*(\bm X+\bm R)}
\label{eq:xiu}
\end{equation}
and its Fourier transform over the sky, averaged over rings of the sky wavenumber $K$, is the directional spectrum
\begin{equation}
P_u(K)=2\pi\int_0^\infty R\,dR\,J_0(KR)\left[\xi_u(R)-\xi_u(\infty)\right]
\label{eq:hankel}
\end{equation}
where $J_0$ is the Bessel function of order zero.

The emissivity has the correlation function
\begin{equation}
\xi_i(R,\Delta z)=\avg{P_i(\bm X,z+\Delta z)\,P_i^*(\bm X+\bm R,z)}
\label{eq:xii}
\end{equation}
and the correlation length $r_i$. The RM density has the mean $\bar\phi$, the rms fluctuation $\sigma_\phi$, the correlation length $r_\phi$ and the spectrum $P_\phi(k)\propto k^{-3-m_\phi}$. Turbulent Faraday rotation turns the polarization by about one radian over the LOS distance
\begin{equation}
\ell=\big(\sqrt2\,\lambda^2\sigma_\phi\big)^{-1}
\label{eq:ell}
\end{equation}
and mean Faraday rotation over the distance
\begin{equation}
\Lb=\big(\lambda^2|\bar\phi|\big)^{-1}
\label{eq:Lb}
\end{equation}
which are the distances $\mathcal L_{\sigma_\phi}$ and $\mathcal L_{\bar\phi}$ of \LP. We label the wavelength by the Faraday rotation strength $\eta=2\lambda^2\sigma_\Phi$, twice the rms angle by which the whole depth rotates the polarization, where $\sigma_\Phi$ is the rms of the RM map. Faraday rotation is weak for $\eta\ll1$, and it is strong and turbulent when $\ell$ is shorter than both correlation lengths.

\subsection{Weak Faraday Rotation}
\label{subsec:weak}

\begin{figure*}[t]
\centering
\includegraphics[width=0.77\textwidth]{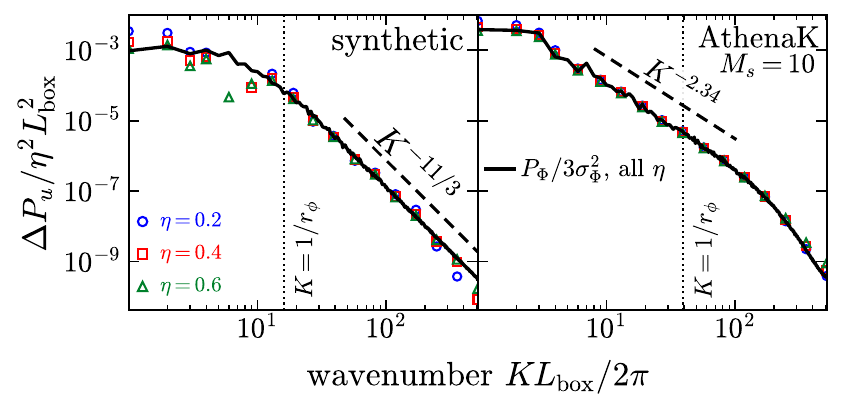}
\caption{Faraday term of the directional spectrum at weak rotation, the change $\Delta P_u=P_u(K,\lambda)-P_u(K,0)$ divided by $\eta^2$, in the synthetic cube with the uniform source (left) and in the supersonic MHD cube (right), for the three values of $\eta$ given by the markers. The solid line is the prediction of Equations~(\ref{eq:weakspec}) and (\ref{eq:fW}) with $f_W=1/3$, computed from the RM map of the same cube. Dashed segments show the slope $K^{-3-m_\phi}$ of the RM spectrum. Wavenumbers are in units of the box, and the dotted line marks $K=1/r_\phi$.}
\label{fig:weak}
\end{figure*}

When the rotation is small on every line of sight, the polarization angle is $\chi=\chi_0+\lambda^2q$ to first order, where $\chi_0$ is the intrinsic polarization angle and
\begin{equation}
q(\bm X)=\int_0^L ds\,W(\bm X,s)\,\phi(\bm X,s)
\label{eq:q}
\end{equation}
is the RM weighted by the fraction $W$ of the emission that originates behind the depth $s$. The directional spectrum then gains the Faraday term (\LPM, Equation~(A9))
\begin{equation}
P_u(K,\lambda)-P_u(K,0)=4\lambda^4P_q(K)
\label{eq:weakspec}
\end{equation}
where $P_q$ is the spectrum of $q$. In a volume much deeper than $r_\phi$ this spectrum is
\begin{equation}
P_q(K)=f_W\,P_\Phi(K)
\label{eq:fW}
\end{equation}
where $P_\Phi(K)\propto K^{-3-m_\phi}$ is the spectrum of the RM map and
\begin{equation}
f_W=\frac{1}{L}\int_0^L W^2\,ds
\label{eq:fWdef}
\end{equation}
is the mean squared weight. A foreground screen has $f_W=1$. Placing the emission inside the rotating volume therefore keeps the slope of the Faraday term and lowers its amplitude, to $f_W=1/3$ for a source that is uniform along the LOS, because the emission from the far side is rotated by the whole depth and the emission from the near side by almost nothing (Figure~\ref{fig:weak}).

\subsection{Strong Turbulent Faraday Rotation}
\label{subsec:strongtheory}

\begin{figure*}[t]
\centering
\includegraphics[width=0.98\textwidth]{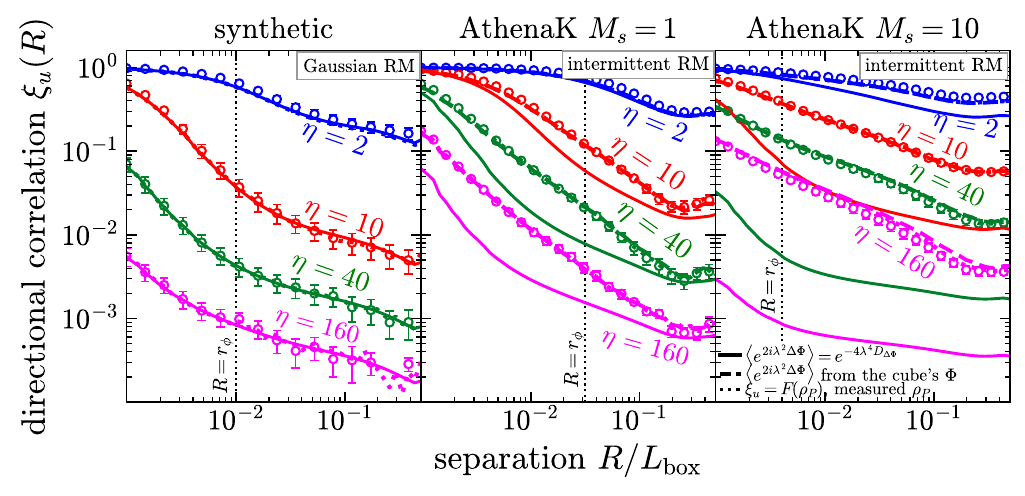}
\caption{Directional correlation at strong rotation in the synthetic cube with the turbulent source (Gaussian RM) and in the two MHD cubes (intermittent RM), LOS $\perp\bm B_0$, for the four values of $\eta$ labeled on the curves. Markers are measured on the Stokes maps, with error bars from the scatter between the eight observer positions. Solid lines are $\xi_u=\mathcal{F}(\rho_P)$, Equation~(\ref{eq:closure}), with $\rho_P$ from the Gaussian average of the Faraday phase factor, Equations~(\ref{eq:xiP}) and (\ref{eq:gaussRM}). Dashed lines (MHD) are $\mathcal{F}(\rho_P)$ with $\rho_P$ from the phase factor averaged over the RMs of the cube, Equation~(\ref{eq:xiPmeas}). Dotted lines are $\mathcal{F}$ applied to the coherence measured on the Stokes maps. For the synthetic cube the two averages coincide and only the solid line is drawn. Separations are in units of the box, and the vertical dotted line marks $R=r_\phi$.}
\label{fig:strong}
\end{figure*}

For an emissivity independent of the RM density and Gaussian fluctuations of the RM density, the correlation of the polarization is (\LPM, Equation~(15))
\begin{equation}
\begin{aligned}
\xi_P(R)=\int_0^L\!dz_1\!\int_0^L\!dz_2\;&\xi_i(R,\Delta z)\,e^{2i\lambda^2\bar\phi\,\Delta z}\\
\times\,&e^{-4\lambda^4D_{\Delta\Phi}(R,z_1,z_2)}
\end{aligned}
\label{eq:xiP}
\end{equation}
where $\Delta z=z_1-z_2$ and $D_{\Delta\Phi}$ is half the variance of the difference $\Delta\Phi$ between the fluctuating RMs of the two emitting points. Its normalized form is the ordinary coherence of the polarization,
\begin{equation}
\rho_P(R)=\frac{\xi_P(R)}{\xi_P(0)} \, .
\label{eq:rhoP}
\end{equation}
The last factor is the average Faraday phase factor of a Gaussian $\Delta\Phi$,
\begin{equation}
\avg{e^{2i\lambda^2\Delta\Phi}}=e^{-4\lambda^4D_{\Delta\Phi}} \, .
\label{eq:gaussRM}
\end{equation}
For any other distribution of $\Delta\Phi$, Equation~(\ref{eq:xiP}) holds with the average $\avg{e^{2i\lambda^2\Delta\Phi}}$ in place of this exponential. We call the two evaluations Gaussian and measured RM statistics, and Figure~\ref{fig:strong} compares them with the simulations (Section~\ref{sec:strong}).

When $\ell$ is much shorter than $r_i$, $r_\phi$ and $L$, only emission from depths within about $\ell$ of each other adds coherently, and the correlation of two lines of sight is proportional to the depth over which their Faraday rotations stay alike, the coherence depth (the effective Faraday depth $W_\phi$ of \LP). Its dependence on the separation sets the break wavenumbers of the directional spectrum. Below the core radius
\begin{equation}
R_c=\left(\frac{\ell^2r_\phi^{\widetilde m_\phi}}{L}\right)^{\frac{1}{1+\widetilde m_\phi}}
\label{eq:Rc}
\end{equation}
where $\widetilde m_\phi=\min(m_\phi,1)$, the two lines of sight rotate alike over the whole depth, so the core wavenumber $K_c=1/R_c$ separates this coherent core from the decorrelated range. At larger separations the coherence depth shrinks, until it equals the separation itself at
\begin{equation}
R_a=r_\phi\left(\frac{\ell}{r_\phi}\right)^{\frac{2}{2+m_\phi}}
\label{eq:Ra}
\end{equation}
which defines $K_a=1/R_a$. For a steep RM spectrum, $1<m_\phi<2$, a third break
\begin{equation}
K_b=\frac{1}{\ell}
\label{eq:Kb}
\end{equation}
marks where the coherence depth becomes shorter than $r_\phi$. At $K<1/r_\phi$ the RM densities of two lines of sight are independent and the spectrum is flat. Between the breaks the directional spectrum has limiting power laws (\LPM, Table~1), for $0<m_\phi<1$ between $K_a$ and $K_c$
\begin{equation}
P_u(K)\propto K^{-1+m_\phi}
\label{eq:inner}
\end{equation}
and a spectrum reaches them only if the neighbouring breaks are several decades apart.

The directions follow from the pair of polarizations at the two positions. For a Gaussian emissivity, this pair is Gaussian at fixed RMs along the two lines of sight, and \LPM\ (their Equations~(6) and (7)) show that $\xi_u$ is the average over these RMs of the directional correlation of a Gaussian pair,
\begin{equation}
\mathcal{F}(\rho)=\frac{\pi}{4}\;{}_2F_1\!\left(\tfrac12,\tfrac12;2;|\rho|^2\right)\rho
\label{eq:F}
\end{equation}
taken at the coherence of the pair at fixed RMs. Here ${}_2F_1$ is the Gauss hypergeometric function. This relation does not assume a Gaussian RM. In a volume that holds many independent depths $\ell$ along each line of sight, the coherence varies little between realizations of the RMs, and
\begin{equation}
\xi_u(R)=\mathcal{F}\big(\rho_P(R)\big)
\label{eq:closure}
\end{equation}
which is the form we evaluate. Near coherence, part of the decorrelation of the polarization is a rotation of one integrated polarization relative to the other, which changes the directions less than in a Gaussian pair, and the exact relation gives a decorrelation $1-\xi_u$ smaller than Equation~(\ref{eq:closure}) by a factor between one and three.

\subsection{Strong Mean Faraday Rotation}
\label{subsec:meantheory}

When the mean RM density dominates and its fluctuations are negligible, the phase factor of Equation~(\ref{eq:xiP}) has unit modulus and the depth integral becomes a Fourier transform along the LOS (\LPM, Equation~(27)). For $L\gg\Lb$
\begin{equation}
\begin{aligned}
\xi_P(R)=\;&L\,\widehat\xi_i(R,\kp)\\
&+\frac{2}{\kp^2}\left[\xi_i(R,0)-\cos(\kp L)\,\xi_i(R,L)\right]
\end{aligned}
\label{eq:fs}
\end{equation}
where the LOS wavenumber selected by the mean rotation is
\begin{equation}
\kp=2\lambda^2\bar\phi
\label{eq:kpar}
\end{equation}
and
\begin{equation}
\widehat\xi_i(R,\kp)=\int_{-\infty}^{\infty}ds\,e^{i\kp s}\,\xi_i(R,s)
\label{eq:xihat}
\end{equation}
is the transverse correlation of the LOS Fourier mode $\kp$ of the emissivity. The second term comes from the two faces of the slab. The wavelength thus selects one LOS wavenumber of the polarized emissivity, and the directional spectrum shows the transverse structure of this mode. For an isotropic source whose correlation along the LOS falls as $\xi_i(0,0)-\xi_i(0,s)\propto|s|^m$, the coherence is (\LPM, Equation~(30))
\begin{equation}
\rho_P(R)=\frac{2^{1-\nu}}{\Gamma(\nu)}\,\big(|\kp|R\big)^\nu\,\mathcal K_\nu\big(|\kp|R\big)
\label{eq:matern}
\end{equation}
where $\nu=(1+m)/2$ and $\mathcal K_\nu$ is the modified Bessel function. The directions then decorrelate on the transverse scale $1/|\kp|$, which shrinks as $\lambda^{-2}$, and at $K\gg|\kp|$ the directional spectrum slowly approaches $K^{-3-m}\ln(K\Lb)$ (\LPM, Equation~(31)). The RMs are the same on all lines of sight, so Equation~(\ref{eq:closure}) is exact for a Gaussian emissivity.

\section{Simulations and Synthetic Observations}
\label{sec:numerics}

\subsection{MHD Simulations}
\label{subsec:athena}

\begin{table*}[t]
\centering\normalsize
\caption{Parameters of the Data Cubes and Fields}
\label{tab:cubes}
\renewcommand{\arraystretch}{1.1}
\begin{tabular*}{\textwidth}{@{\extracolsep{\fill}}lccccccc}
\hline\hline
Data & Grid & LOS & $m_\phi$ & $L/r_\phi$ & $L/r_i$ & $|\avg{P_i}|^2/\avg{|P_i|^2}$ & Figures \\
\hline
MHD, $M_A=0.8$, $M_s=1$ & $1024^3$ & $\perp B_0$ & $0.37$ & 32 & 37 & 0.47 & \ref{fig:overview}, \ref{fig:strong} \\
MHD, $M_A=0.8$, $M_s=10$ & $1024^3$ & $\perp B_0$ & $-0.66$ & 250 & 35 & 0.68 & \ref{fig:overview}--\ref{fig:strong} \\
MHD, $M_A=0.8$, $M_s=1$ & $1024^3$ & $\parallel B_0$ & & & 44 & 0.004 & \ref{fig:mean} \\
MHD, $M_A=0.8$, $M_s=10$ & $1024^3$ & $\parallel B_0$ & & & 45 & 0.006 & \ref{fig:mean} \\
Synthetic cube, turbulent source & $1024^3$ & & $2/3$ & 100 & 11 & 0.20 & \ref{fig:overview}, \ref{fig:strong}, \ref{fig:mean} \\
Synthetic cube, uniform source & $1024^3$ & & $2/3$ & 100 & uniform & 0.70 & \ref{fig:weak} \\
High-resolution Gaussian RM, constant source & Table~\ref{tab:gauss} & & $2/3$ & 10 & uniform & 1 & \ref{fig:regimes} \\
High-resolution Gaussian RM, constant source & Table~\ref{tab:gauss} & & $3/2$ & 30 & uniform & 1 & \ref{fig:regimes} \\
\hline
\end{tabular*}
\vspace{4pt}

{\raggedright \textit{Note.} The high-resolution Gaussian RM fields have their own grids, given in Table~\ref{tab:gauss}. For the cubes, $r_\phi$ is the correlation length of the RM density and $r_i$ that of the emissivity along the LOS, the separations at which the correlations fall to one half. The index $m_\phi$ is that of the spectrum of the RM density, $k^{-3-m_\phi}$. The sixth column is the fraction of the variance of the intrinsic polarization that is in its mean. For pure mean rotation the sources are combined with a constant RM density.\par}
\end{table*}

The MHD data are snapshots of the two isothermal \textsc{AthenaK} \citep{Stone2026} simulations of \MLP, on a $1024^3$ grid in the saturated state. Both have the Alfv\'en Mach number $M_A=0.8$, so the magnetic field $\bm B$ fluctuates around a dynamically important mean field $\bm B_0$, and their sonic Mach numbers are $M_s=1$ and $M_s=10$ (Table~\ref{tab:cubes}). Each element emits with $P_i\propto(B_1+iB_2)^2$, where $B_1$ and $B_2$ are the field components in the plane of the sky. This is the synchrotron emissivity of relativistic electrons with a power-law energy spectrum of index 3, uniform in space, for which the polarized intensity is proportional to the square of the magnetic field in the plane of the sky \citep{LazarianPogosyan2012}. Each element also rotates with $\phi\propto n_eB_\parallel$, where $B_\parallel$ is the field component along the LOS and $n_e$ is proportional to the density.

For weak and strong turbulent rotation the LOS is perpendicular to $\bm B_0$, so the mean field lies in the plane of the sky and gives every line of sight a net polarization, and the RM density has zero mean. In the supersonic run shocks give the density a shallow spectrum \citep{Beresnyak2005, Kowal2007}, and the RM density, dominated by the density fluctuations, has a spectrum shallower than $k^{-3}$. For the Stokes maps of Figures~\ref{fig:overview} and \ref{fig:strong} we shift the source by half the cube across the sky relative to the RM density, which removes any correlation between emission and rotation, as Equation~(\ref{eq:xiP}) assumes, and changes the results little. For strong mean rotation the LOS is parallel to $\bm B_0$ (Section~\ref{sec:mean}).

\subsection{Synthetic Fields}
\label{subsec:synthetic}

In the synthetic cubes, of depth $L=100\,r_\phi$, the RM density is a Gaussian random field with zero mean and the spectrum $(k^2+r_0^{-2})^{-11/6}$, which gives $m_\phi=2/3$ below the injection scale $r_0$. We consider two cases. In the first case the source is turbulent and is made from a synthetic magnetic field whose two components in the plane of the sky are $B_1=1+0.6\,g_1$ and $B_2=0.6\,g_2$, where $g_1$ and $g_2$ are independent Gaussian random fields of unit variance with the Kolmogorov spectrum, and it emits with $P_i\propto(B_1+iB_2)^2$, as in the MHD cubes. This magnetic field is a uniform mean field in the plane of the sky with Gaussian Kolmogorov fluctuations, so in the synthetic cubes the Gaussian fields are the RM density and the fluctuations of the magnetic field. The emissivity, quadratic in the magnetic field, is not Gaussian and has a mean polarization, mimicking synchrotron emission. In the second case the source is uniform, with an emissivity that is constant along each line of sight and an angle that varies only across the sky, while the RM density is the same Gaussian field as in the first case. The second case is used for the weak-rotation test of Figure~\ref{fig:weak}. In both cases the source and the RM density are independent, and for strong mean rotation the turbulent source is combined with a constant RM density in place of the Gaussian one. The regimes of strong turbulent rotation require $\ell\ll r_\phi\ll L$ and a resolved coherent core, which none of the cubes reaches. For these regimes we generate Stokes maps and pairs of lines of sight directly at the required resolution, on grids separate from the $1024^3$ cubes, with the realizations listed in Table~\ref{tab:gauss} and described in Appendix~\ref{app:gauss}. This is possible in the special case of a constant source, with the same polarized emissivity everywhere, for which the direction on a line of sight depends only on the RM density along it. We combine it with Gaussian RM densities with $m_\phi=2/3$ and $m_\phi=3/2$, on either side of $m_\phi=1$, where the order of the breaks changes. Figure~\ref{fig:regimes} shows how direct calculations with pairs of lines of sight extend the range of the spectra for $\ell=0.01\,r_\phi$, where the coherent core is not resolved by the Stokes maps, while for $\ell=0.1\,r_\phi$ the Stokes maps alone are sufficient at all wavenumbers. In both cases the measured spectra follow the prediction through all break wavenumbers. The limiting power laws between the breaks, however, are not reached, because the breaks are only one to two decades apart, so the spectrum shows the breaks rather than the limiting slopes (Section~\ref{sec:regimes}).

\subsection{Synthetic Observations and Predictions}
\label{subsec:synobs}

The cubes have $1024^3$ cells, so each Stokes map has $1024^2$ lines of sight, and each line of sight is sampled with the same spacing as the lines of sight on the sky. Equation~(\ref{eq:Pmixed}) is integrated along the LOS for wavelengths that cover $\eta$ from 0.05 to 160. At strong rotation only the emission within a few $\ell$ of the observer contributes, so we make the Stokes maps for eight observer positions along the periodic LOS and average their statistics. From each pair of Stokes maps we compute $\xi_u(\bm R)$ and $\xi_P(\bm R)$, average them over directions on the sky, and obtain the ring-averaged spectrum $P_u(K)$.

The predictions use the three-dimensional correlations of $P_i$ and $\phi$ measured in the same cube. With Gaussian RM statistics, Equation~(\ref{eq:xiP}) is evaluated with $\xi_i$ and the structure function of the RMs. With measured RM statistics, the average over the periodic cube turns Equation~(\ref{eq:xiP}) into
\begin{equation}
\xi_P(R)=L\int_0^L d\Delta z\,\xi_i(R,\Delta z)\,C_\zeta(R,\Delta z)
\label{eq:xiPmeas}
\end{equation}
where $C_\zeta$ is the three-dimensional autocorrelation of the Faraday phase factor $e^{2i\lambda^2\Phi(\bm X,z)}$, computed from the RMs of the cube without any assumption about their distribution. In both cases $\rho_P$ is converted to $\xi_u$ with Equation~(\ref{eq:closure}) and transformed in the same way as the Stokes maps.

\section{Results}
\label{sec:results}

\begin{figure*}[t]
\centering
\includegraphics[width=0.98\textwidth]{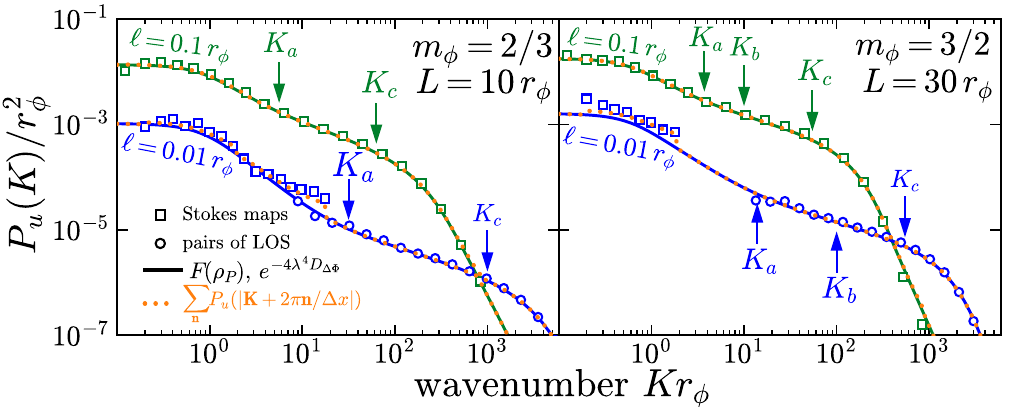}
\caption{Directional spectrum through the regimes of strong turbulent rotation in the high-resolution Gaussian RM fields of Table~\ref{tab:gauss}, with a constant source and a Gaussian RM density with $m_\phi=2/3$ (left) and $m_\phi=3/2$ (right), at two wavelengths, $\ell=0.1\,r_\phi$ (green) and $\ell=0.01\,r_\phi$ (blue). Solid lines are $\mathcal{F}(\rho_P)$, Equation~(\ref{eq:closure}), with $\rho_P$ evaluated exactly for the simulated field. Squares are Stokes maps, from large cubes at low $K$ and small cubes at high $K$, and circles are the transform of $\xi_u$ measured on simulated pairs of lines of sight (Appendix~\ref{app:gauss}). For $\ell=0.1\,r_\phi$ the maps resolve the coherent core, and the squares cover the whole range. For $\ell=0.01\,r_\phi$ the core is far smaller than the spacing of any feasible map, and the pairs continue the spectrum to high $K$. Dotted lines are the prediction sampled on the lattice of lines of sight of the Stokes maps (Appendix~\ref{app:gauss}), drawn over the solid lines, from which they depart only near the resolution limit of the maps. Arrows mark the break wavenumbers of Equations~(\ref{eq:Rc}), (\ref{eq:Ra}) and (\ref{eq:Kb}) in the color of each curve, the core wavenumber $K_c$, beyond which the two lines of sight rotate alike over the whole depth, $K_a$, where the coherence depth equals the separation, and for $m_\phi=3/2$ $K_b$, where it equals the correlation length $r_\phi$ of the RM density.}
\label{fig:regimes}
\end{figure*}

\begin{figure*}[t]
\centering
\includegraphics[width=0.98\textwidth]{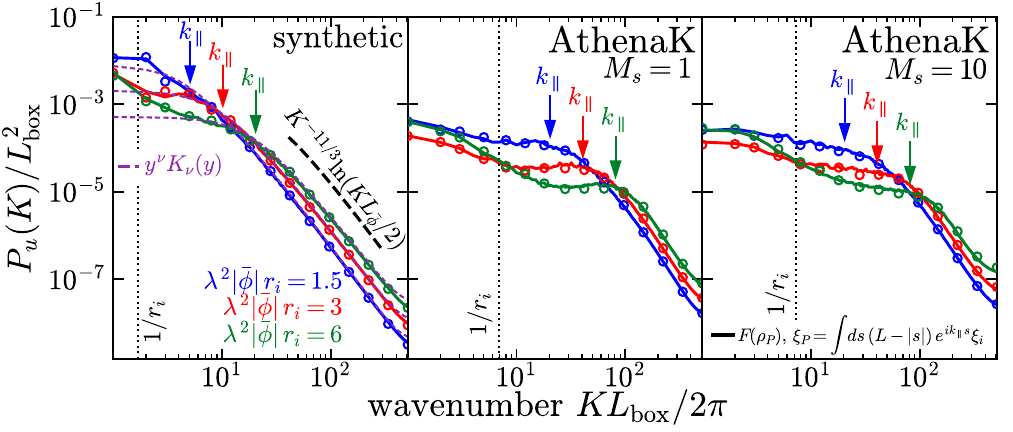}
\caption{Directional spectrum under pure mean Faraday rotation of the sources of the two MHD cubes (LOS $\parallel\bm B_0$) and of the synthetic turbulent source, for the three wavelengths labeled in the left panel by $\lambda^2|\bar\phi|\,r_i$, the mean rotation in radians over one correlation length of the source. Markers are measured on the Stokes maps. Solid lines are $\mathcal{F}(\rho_P)$, Equation~(\ref{eq:closure}), with $\xi_P$ of Equation~(\ref{eq:fs}) computed from the source correlation of the cube. Purple dashed lines (synthetic) use the closed form for an isotropic source, Equation~(\ref{eq:matern}), which omits the mean polarization of the source and therefore falls below at the lowest $K$. The black dashed line shows the high-$K$ form $K^{-11/3}\ln(K\Lb/2)$. Colored arrows mark the selected LOS wavenumber $\kp=2\lambda^2|\bar\phi|$ of each wavelength. Wavenumbers are in units of the box, and the dotted line marks $K=1/r_i$.}
\label{fig:mean}
\end{figure*}

\subsection{From Weak to Strong Faraday Rotation}
\label{sec:overview}

Without Faraday rotation the directional spectrum is that of the intrinsic polarization directions (Figure~\ref{fig:overview}, black). In the MHD cubes it is steep, because the mean magnetic field in the plane of the sky keeps the directions ordered, and it drops further in the dissipation range. Weak Faraday rotation adds a term that grows as $\lambda^4$ and has the slope of the RM map. In the supersonic cube the RM density, shaped by shocks, has far more small-scale structure than the intrinsic directions. Already at $\eta=0.2$, an rms rotation of a tenth of a radian, the Faraday term dominates the spectrum, by two orders of magnitude at the smallest scales, and a single pair of Stokes maps shows the RM spectrum rather than the magnetic field in the plane of the sky. In the transonic cube and in the synthetic cube the two terms have similar slopes, and rotation raises the amplitude without changing the slope. The predicted sum follows the Stokes maps in all three cubes, also in the dissipation range, because it is computed from the RM map itself rather than from a power law.

Once the rotation through the volume exceeds a radian, distant lines of sight decorrelate, and the spectrum loses power at large scales and flattens. The predictions follow the Stokes maps through this transition, for the MHD cubes only when the measured RM statistics are used, for the reason shown in Section~\ref{sec:strong}. At $\eta=2$ neighboring lines of sight are still nearly coherent, and the predictions lie above the maps at small scales, as expected near coherence (Section~\ref{subsec:strongtheory}). In the synthetic cube at strong rotation the spectrum follows the flat limiting slope $K^{-1/3}$ of Equation~(\ref{eq:inner}) from $K_a$ to the resolution limit. The MHD cubes do not show this range, because their coherent core is not resolved, and in the supersonic run the RM spectrum lies outside the range of Equation~(\ref{eq:inner}).

\subsection{The Faraday Term at Weak Rotation}
\label{sec:weak}

Subtracting the spectrum without rotation isolates the Faraday term (Figure~\ref{fig:weak}). Divided by $\eta^2$, Equations~(\ref{eq:weakspec}) and (\ref{eq:fW}) predict one curve for all wavelengths, which we compute from the RM map of the same cube with $f_W=1/3$. The measured terms collapse onto this curve, in the synthetic cube with the uniform source and in the supersonic MHD cube over the whole range of scales, including the dissipation range, and the transonic cube behaves in the same way. The emissivity of the MHD cubes fluctuates, but the mean magnetic field in the plane of the sky dominates the emission, so their weight $W$ is close to that of a uniform source. The slope is that of the RM map, so the Faraday term keeps the slope it has behind a screen, and only its amplitude is lower. At the lowest wavenumbers the synthetic cube falls below the curve, because there the intrinsic directions of two lines of sight differ by a large angle, which reduces the Faraday term and which Equation~(\ref{eq:weakspec}) neglects.

\subsection{Strong Turbulent Rotation and Intermittency}
\label{sec:strong}

The directional correlation shows how strong rotation acts on the MHD cubes (Figure~\ref{fig:strong}). With Gaussian RM statistics the predicted correlation lies below the Stokes maps, and the gap grows with the rotation strength and with the Mach number, up to a factor of about 4 in the transonic cube and about 50 in the supersonic cube. With the RMs of the cube, Equation~(\ref{eq:xiPmeas}), the same relations follow the Stokes maps at every wavelength. The steps from the RMs to the coherence of the polarization, and from the coherence to the directions, therefore hold in MHD turbulence, and what fails is the Gaussian average of the Faraday phase factor, Equation~(\ref{eq:gaussRM}).

That average describes the RM differences between lines of sight by their variance alone. The RM of MHD turbulence is intermittent. Most neighboring lines of sight differ little in RM, while a few, which cross shocks and current sheets, differ strongly. The rare large differences set the variance, so the Gaussian average decorrelates all pairs while most of them remain coherent. The RM maps confirm this picture, with large RM differences much more frequent than for a Gaussian distribution and with a measured phase-factor average that exceeds its Gaussian value by two orders of magnitude in the supersonic cube at $\eta=10$. Such statistics are expected from the density structure of supersonic turbulence \citep{Federrath2008, Seta2021, Seta2023}, which is why the departure grows with the Mach number. The shifted source excludes a correlation between emission and rotation as the cause.

In the synthetic cube the RM density is Gaussian and independent of the source, as Equation~(\ref{eq:xiP}) assumes, the two evaluations coincide, and $\mathcal{F}(\rho_P)$ follows the measured correlation at all separations and wavelengths. Applied to the coherence measured on the Stokes maps, Equation~(\ref{eq:closure}) reproduces $\xi_u$ in all three cubes (dotted lines), so the conversion from coherence to directions holds also for intermittent RM fluctuations. The only departure is near coherence, at $\eta=2$ and small separations, where the directions stay more correlated than $\mathcal{F}(\rho_P)$ gives, by a factor of up to about two, within the range from one to three that the exact relation allows. This is the departure of the $\eta=2$ spectra in Figure~\ref{fig:overview}.

\subsection{Breaks of the Directional Spectrum}
\label{sec:regimes}

The MHD cubes neither resolve the coherent core nor span enough decades to separate the break wavenumbers. Gaussian RM fields satisfy Equation~(\ref{eq:xiP}) exactly and can be sampled at any separation (Figure~\ref{fig:regimes}). Their directional spectrum has the shape that Section~\ref{subsec:strongtheory} describes. It is flat at $K<1/r_\phi$, where the RMs of two lines of sight are independent. It steepens up to $K_a$, flattens between $K_a$ and $K_c$, and falls steeply beyond $K_c$, inside the coherent core. As the wavelength grows, all breaks move to smaller scales, and the spectrum drops as the directions decorrelate.

The Stokes maps follow the prediction across all breaks where they resolve the core, and the simulated pairs follow it where the core is far below the spacing of the lines of sight in any feasible map. The spectra bend at the predicted break wavenumbers but do not reach the limiting slopes between them, including the intermediate branch of $m_\phi=3/2$, because the breaks are only one to two decades apart. Observed directional spectra will likewise show the breaks, and their motion with wavelength, more readily than the limiting slopes, and their interpretation needs the full prediction rather than the asymptotic power laws.

\subsection{Fourier Selection by Mean Rotation}
\label{sec:mean}

Along the mean field the RM density of the MHD cubes has fluctuations comparable to its mean, so turbulent rotation is not negligible. To isolate the regime of Section~\ref{subsec:meantheory} we keep the sources of the MHD cubes viewed along $\bm B_0$, and the synthetic turbulent source, and replace the RM density by a constant one.

Equation~(\ref{eq:fs}), evaluated with the source correlation of the same cube, follows the Stokes maps of all three sources at all three wavelengths (Figure~\ref{fig:mean}). As the wavelength grows, the selected LOS wavenumber $\kp$ increases, the directions decorrelate over shorter transverse distances, and power moves from large to small scales. The spectra of the MHD sources are nearly flat at $K<\kp$ and fall steeply beyond, so that each wavelength produces a break near $K=\kp$ that moves to higher $K$ as $\lambda^2$ increases. A sequence of wavelengths thus scans the emissivity spectrum along the LOS. In the MHD cubes this spectrum is anisotropic and reaches the dissipation range at the shortest scales, where the isotropic closed form, Equation~(\ref{eq:matern}), does not apply but Equation~(\ref{eq:fs}) does. For the isotropic synthetic source the closed form gives the full prediction except at the lowest $K$, and at high $K$ the spectrum follows the logarithmic form of \LPM\ over nearly one decade, shallower than $K^{-11/3}$.

\section{Discussion}
\label{sec:discussion}

The relations of Section~\ref{subsec:strongtheory} separate into a part that holds for any RM statistics, Equation~(\ref{eq:xiP}) with the measured phase-factor average and the passage from the coherence to the directions, and the Gaussian average of the Faraday phase factor, which fails in MHD turbulence. The possibility that the intermittent density of supersonic turbulence weakens the Gaussian damping of the directional correlation was raised for the screen (\MLP). For coincident regions the effect exceeds an order of magnitude. At strong rotation the directional correlation therefore depends on the whole distribution of the RM differences, and its excess over the Gaussian value measures the intermittency of the magneto-ionic medium \citep[compare][]{Wang2024}. The polarization variance $\avg{|P|^2}$ of the supersonic cube at the longest wavelength also exceeds its Gaussian prediction, by nearly an order of magnitude, so a Gaussian model fitted to the depolarization or to the directional correlation of such a medium would underestimate its RM dispersion.

Coincident regions differ from a screen mainly at strong rotation. A screen randomizes the directions without changing the polarized amplitude, whereas in coincident regions the emission from different depths cancels, and the directions stay correlated over a scale that shrinks with wavelength. At weak rotation the two geometries differ only by the amplitude $f_W$, and the Faraday term can be read from a single pair of Stokes maps wherever it exceeds the intrinsic spectrum, as at the small scales of our MHD cubes.

The predictions were evaluated with three-dimensional statistics that an observer does not have, and applying them to data requires the inverse step, in which the wavelength dependence of the directional statistics constrains them. Viewed along the mean magnetic field, turbulent and mean rotation act together in the MHD cubes, and the combined case, closest to observations of diffuse Galactic emission along the field, needs a separate study. Noise and beam smoothing, absent from our Stokes maps, affect the normalization of weakly polarized lines of sight and must be included before applications to surveys with the Square Kilometre Array and its precursors \citep{Heald2020}.

\section{Summary}
\label{sec:summary}

We tested the theory of \LPM\ for the polarization directions of a turbulent volume that both emits and Faraday-rotates, with synthetic observations of MHD simulations and Gaussian random fields. The simulations agree with the predictions in all three regimes of Faraday rotation, provided that the RM fluctuations of MHD turbulence keep their actual statistics. Our main conclusions are as follows.

\begin{enumerate}
\item At short wavelengths Faraday rotation adds to the directional spectrum a term with the slope of the RM spectrum, as behind a screen, but with one third of the screen amplitude for uniform emission. In MHD turbulence this term dominates the small scales already when the rotation is a fraction of a radian, so a single polarization map measures the RM spectrum at small scales.

\item At long wavelengths the relations of the theory hold in MHD turbulence once the RM differences keep their actual distribution. A Gaussian RM model underestimates the directional correlation, by more than an order of magnitude in supersonic turbulence, because most neighboring lines of sight differ little in RM while a few differ strongly. The directional correlation is therefore a probe of the intermittency of the magneto-ionic medium.

\item The directional spectrum bends at the predicted break wavenumbers, but its limiting slopes require separations of scales of several decades, beyond those simulated here.

\item Under strong mean rotation the wavelength selects one LOS Fourier mode of the polarized emissivity, so a sequence of wavelengths scans the emissivity spectrum along the LOS, also for anisotropic MHD turbulence.
\end{enumerate}

\section*{Acknowledgments}
A.L. acknowledges the support of NSF grant AST 2307840.

\appendix

\section{Gaussian RM Fields at High Resolution}
\label{app:gauss}

\begin{table}[h]
\centering\normalsize
\caption{Realizations of the High-resolution Gaussian RM Fields}
\label{tab:gauss}
\renewcommand{\arraystretch}{1.1}
\begin{tabular}{cccccc}
\hline\hline
$m_\phi$ & $\ell/r_\phi$ & Cubes & Size & $\Delta x/r_\phi$ & Generation \\
\hline
$2/3$ & 0.1 & 4 & $6300^2\times300$ & 0.02 & grid \\
$2/3$ & 0.1 & 4 & $6720^2\times300$ & 0.003 & exact \\
$3/2$ & 0.1 & 3 & $6300^2\times900$ & 0.02 & grid \\
$3/2$ & 0.1 & 2 & $6720^2\times900$ & 0.003 & exact \\
$2/3$ & 0.01 & 4 & $2000^2\times3000$ & 0.03 & exact \\
$2/3$ & 0.01 & 1 & $1600^2\times3000$ & 0.005 & exact \\
$3/2$ & 0.01 & 2 & $1600^2\times9000$ & 0.04 & exact \\
\hline
\end{tabular}
\vspace{4pt}

{\raggedright \textit{Note.} The size is the number of lines of sight times the number of samples along each of them within the depth $L$. The lines of sight are spaced by $\Delta x$ on the sky, and the field is sampled along them at intervals of $\ell/3$. The cubes are periodic.\par}
\end{table}

The Gaussian RM densities of Figure~\ref{fig:regimes} have zero mean and the spectrum $(k^2+r_0^{-2})^{-(3+m_\phi)/2}$, with $L=10\,r_\phi$ for $m_\phi=2/3$ and $L=30\,r_\phi$ for $m_\phi=3/2$. The Stokes maps come from the three-dimensional cubes of Table~\ref{tab:gauss}, and each measured spectrum in Figure~\ref{fig:regimes} combines two sets of cubes, a large box with coarse spacing for the low wavenumbers and a small box with fine spacing for the high wavenumbers. Unlike in the MHD and synthetic cubes, the sampling along the LOS, $\ell/3$, differs from the spacing $\Delta x$ of the lines of sight on the sky. For $\ell=0.1\,r_\phi$ the cubes with $\Delta x=0.02\,r_\phi$ give the spectrum at $Kr_\phi<3$, and the cubes with $\Delta x=0.003\,r_\phi$ at higher $K$. The former are generated in the usual way, with Gaussian Fourier amplitudes whose variance follows the spectrum and a Fourier transform to the grid, so they hold the fluctuations of the field only on scales longer than $\Delta x$.

The other cubes are generated so that the values on each line of sight are exact samples of the continuous field. This matters for $\ell=0.01\,r_\phi$, whose core radius, $R_c\approx10^{-3}\,r_\phi$, is far below any spacing that a box of many $r_\phi$ allows. The direction on a line of sight depends only on the RM density along that line, so the field is needed only on the $N^2$ lines. For each LOS wavenumber $k_z$ the transverse correlation of the RM density is $S_0(k_z)f(\kappa R)$, with $\kappa^2=k_z^2+r_0^{-2}$, $S_0\propto\kappa^{-1-m_\phi}$ and $f(x)\propto x^{\nu_\phi}\mathcal K_{\nu_\phi}(x)$, $f(0)=1$, $\nu_\phi=(1+m_\phi)/2$, and we generate each $k_z$ plane from this correlation at the separations of the lines of sight, summed over the periodic images of the box. Where $\Delta x$ is much larger than $R_c$, the directions on neighboring lines of sight are nearly independent and add a white-noise level close to $\Delta x^2$ to the spectrum, which we measure at the high-$K$ end of the maps and subtract.

The pairs of lines of sight at a separation $R$ have RM densities generated along $z$ from their one-dimensional cross-spectrum $S_0(k_z)f(\kappa R)$, so that $R$ can lie decades below $r_\phi$, with $2\times10^6$ to $6\times10^7$ pairs per separation. Their directional correlation is transformed with Equation~(\ref{eq:hankel}), continued by the prediction outside the separations where it exceeds three standard deviations of the mean. The Stokes maps sample $u$ on a lattice of spacing $\Delta x$, so their predicted spectrum includes the images above the Nyquist wavenumber, $P_u^{\rm lat}(\bm K)=\sum_{\bm n}P_u(|\bm K+2\pi\bm n/\Delta x|)$ over pairs of integers $\bm n$.

\clearpage
\onecolumngrid
\begin{center}ORCID iDs\end{center}
\noindent A. Melnichenka\,\includegraphics[width=9pt]{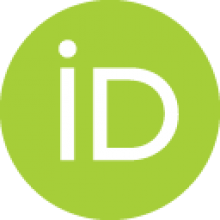} \url{https://orcid.org/0009-0004-1969-779X}\\
A. Lazarian\,\includegraphics[width=9pt]{Orcid-ID.png} \url{https://orcid.org/0000-0002-7336-6674}\\
D. Pogosyan\,\includegraphics[width=9pt]{Orcid-ID.png} \url{https://orcid.org/0000-0002-7998-6823}


\end{document}